# Voice Disorder Detection Using Long Short Term Memory (LSTM) Model

Vibhuti Gupta
Department of Computer Science
Texas Tech University, Lubbock, TX 79415
Email: vibhuti.gupta@ttu.edu

*Abstract*— **Automated detection of voice disorders with computational methods is a recent research area in the medical domain since it requires a rigorous endoscopy for the accurate diagnosis. Efficient screening methods are required for the diagnosis of voice disorders so as to provide timely medical facilities in minimal resources. Detecting Voice disorder using computational methods is a challenging problem since audio data is continuous due to which extracting relevant features and applying machine learning is hard and unreliable. This paper proposes a *Long short term memory model* (LSTM) to detect pathological voice disorders and evaluates its performance in a real 400 testing samples without any labels. Different feature extraction methods are used to provide the best set of features before applying LSTM model for classification. The paper describes the approach and experiments that show promising results with 22% sensitivity, 97% specificity and 56% unweighted average recall.**

*Keywords— Neoplasm; Phonotrauma; Vocal Paralysis; Long Short Term Memory; Mel frequency cepstral coefficient*

I. INTRODUCTION

A voice disorder occurs due to disturbance in respiratory, laryngeal, subglottal vocal tract or physiological imbalance among the system which causes abnormal voice quality, pitch and loudness as compared to normal voice of a healthy person[21]. Major voice disorders include vocal nodules polyps, and cysts (collectively referred as Phonotrauma), glottis neoplasm; and unilateral vocal paralysis. Voice disorders may affect a person's social, professional and personal aspects of communication which hinders its growth in all these aspects [2].

Current approaches for voice disorder detection requires rigorous endoscopy (i.e. laryngeal endoscopy) which is a multistep examination including mirror examination, rigid and flexible laryngoscopy and videostroboscopy [1][22]. This rigorous examination requires a lot of expensive medical resources and delays the diagnosis of voice disorders due to which treatment get delayed which worsen the severity of the disease. Sometimes voice disorders remain unidentified since they are considered as normal by most of the people due to inefficient and slow screening methods. Accuracy in diagnosis is also important to cure the correct disorder with proper treatment.

Automated detection of voice disorders is crucial to mitigate these problems since it makes the diagnosis process simpler, cheaper and less time consuming. Recent research on computerized detection of voice disorders has studied various machine learning techniques and few deep learning techniques [3-5, 6-10, 11-13]. Majority of the previous work deals with machine learning techniques for voice disorder detection [3,4]. [3] used rule based analysis by analyzing various acoustic measures such as Fundamental Frequency, jitter, shimmer etc. and then applied logistic model tree algorithm, instance based learning and SVM algorithms while [4] used SVM and decision trees for detecting voice disorders. Muhammad et. al [5] used gaussian mixture model (*GMM*) to classify 6 different types of voice disorders.

Deep learning is widely used nowadays for image recognition, music genre classification and various other applications. It is recently used for voice disorder detection tasks [6-10]. Most recently [6] applied deep neural networks (*DNN*) for voice disorder detection using dataset of Far Eastern Memorial Hospital (*FEMH*) with 60 normal voice samples and 402 various voice disorder samples and achieved the highest accuracy as compared to other machine learning approaches. Authors at [7] discussed the use of deep neural networks (DNN) in acoustic modeling. They have applied DNN in various speech recognition tasks and found that DNN's are performing well. Wu et. al [8] used convolutional neural network (CNN) for vocal cord paralysis which is a challenging medical classification problem. Alhussein et. al [9] applied deep learning into a mobile healthcare framework to detect voice disorder.

Despite the success of above mentioned models, recurrent neural networks (RNN) are not used for voice disorder tasks. Recurrent neural networks are widely used for speech recognition, music genre classification, natural language processing and sequence prediction problems [11-12] . Long short term memory (LSTM) is a special type of recurrent neural network which is widely used for long term dependencies. [11] used LSTM for voice activity detection which separates the incoming speech with noise. Convolutional neural networks are used along with LSTM in [12] to determine dysarthric speech disorder. To the best of our knowledge, none of the studies used LSTM for voice disorder detection task.

Our major contributions in this paper are: (1) to propose an approach to detect pathological voice disorders using Long short term memory (LSTM) model. (2) to evaluate LSTM performance in differentiating normal and pathological voice samples. The rest of the paper is organized as follows. Section II discusses the material and methods. Section III describes our experimental setup along with results and Section IV concludes the paper.

## II. METHOD

This section provides a brief overview of our proposed approach with the general description of *Long short Term memory Model (LSTM)* used in our experiments and description of dataset with preprocessing part.

### A. Overview of proposed approach

Our proposed approach starts by loading the input voice samples provided by *Far Eastern Memorial Hospital (FEMH)* voice disorder detection challenge [16] as shown in Figure 1 which includes 50 normal voice samples and 150 samples of common voice disorders, including vocal nodules, polyps, and cysts (collectively referred as Phonotrauma); glottis neoplasm; and unilateral vocal paralysis, that comprises of our training dataset.

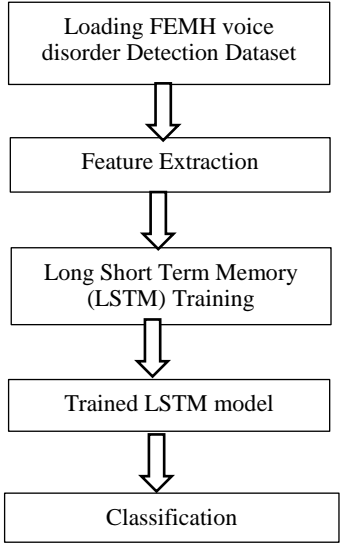

**Fig. 1** Overview of proposed approach

Feature extraction process is done after loading the data that includes Mel-frequency cepstral coefficients (MFCC), spectral centroid, chroma and spectral contrast features comprising 33 features for each audio sample. Details are provided in the further sections. Then LSTM model is used to train the model which is used for classification.

### B. Long Short Term Memory (LSTM) Model

Long Short term Memory Network (LSTM) are the special type of Recurrent Neural networks capable of learning long term dependencies [17]. A typical LSTM network has 4 layers i.e. input layer, 2 hidden layers and one output layer. It contains three gates *forget gate, Input gate* and *Output gate*.

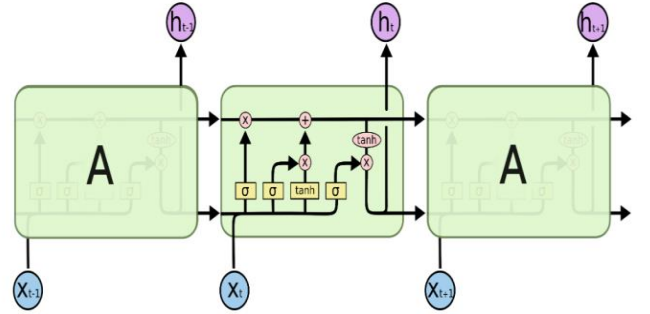

**Fig. 2** LSTM Network

Forget gate layer decides what information has to be kept or thrown away from the cell state. It takes input as $h_{t-1}$ and $x_t$ and outputs a number between 0 and 1 using $f_t$ as in the Eqn(1). Value of 0 indicates completely remove and 1 to completely keep this.

$$f_t = \sigma(W_f [h_{t-1}, x_t] + b_f) \quad (1)$$

Now we need to decide what information has to be stored in the cell state. It has two parts, firstly input gate layer using to decide what values has to be updated and then tanh layer generates a vector of new candidate values that has to be added. $i_t$ is the function used by input gate layer and $C$ is the vector of new candidate values by *tanh* layer as shown in the Eqn (2) and (3).

$$i_t = \sigma(W_i [h_{t-1}, x_t] + b_i) \quad (2)$$

$$C = tanh(W_C [h_{t-1}, x_t] + b_C) \quad (3)$$

Updated state of cell is shown in Eqn (4)

$$C_t = f_t * C_{t-1} + i_t * C \quad (4)$$

Finally, we need to decide what will be the output using output gate. First we run the sigmoid layer using $o_t$ as shown in the Eqn (5) and then its output is multiplied by *tanh* to get the output which is shown in Eqn. (6)

$$o_t = \sigma(W_o [h_{t-1}, x_t] + b_o) \quad (5)$$

$$h_t = o_t * tanh(C_t) \quad (6)$$

$$output_{class} = \sigma(h_t * W_{outparameter}) \quad (7)$$

The output class of the LSTM network is determined by the Eqn. (7) $W_f, W_i, W_C, W_o, W_{outparameter}$ are the weights, $b_f$, $b_i$, $b_C$, $b_o$ are the biases, $h_t$ is the output at time t, $x_t$ are the input features and $output_{class}$ is the classification output.

## C. Dataset and Preprocessing

The dataset comprises of 200 samples in the training set and 400 samples in the testing set. Out of 150 common voice disorder samples, 40 are for *glottis neoplasm*, 60 for *Phonotrauma* and 50 are for *vocal palsy* in the training set. The labels of training dataset includes gender, age, whether the speaker is healthy or not and the corresponding voice disease.

Voice samples of a 3-second sustained vowel sound were recorded at a comfortable level of loudness, with a microphone-to-mouth distance of approximately 15–20 cm, using a high-quality microphone (Model: SM58, SHURE, IL), with a digital amplifier (Model: X2u, SHURE) under a background noise level between 40 and 45 dBA. The sampling rate was 44,100 Hz with a 16-bit resolution, and data were saved in an uncompressed .wav format as used in [6]. Further dataset information is given in [6][16].

Visualization of voice samples is done using the waveforms as shown in Figures 3,4,5,6. It shows waveform whose y-axis represents the amplitude of voice sample and x-axis as time duration. We plotted 4 secs duration of each type of voice sample with a sampling rate of 22050 Hz.

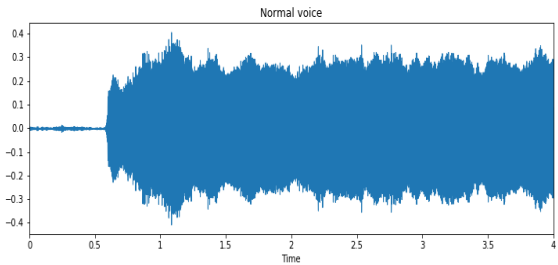

**Fig. 3** Waveform of Normal Voice Sample

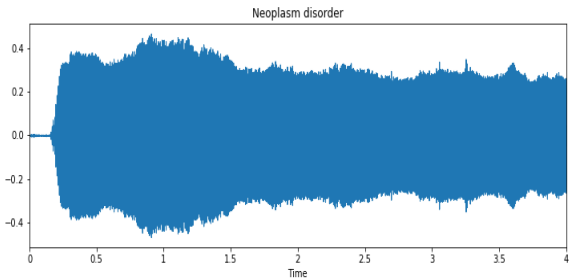

**Fig. 4** Waveform of Neoplasm Voice disorder Sample

As shown in Figures 3 and 4 normal voice sample amplitude is fluctuating while neoplasm disorder waveform not. Waveforms for phonotrauma and vocal palsy as shown in Figures 5 and 6 shows variations as compared to normal and neoplasm voice samples. Amplitudes for Phonotrauma disorder are fluctuating and increasing while for vocal palsy, its decreasing. Normal voice sample can be easily distinguishable with Phonotrauma and Vocal palsy disorder but not much with Neoplasm disorder.

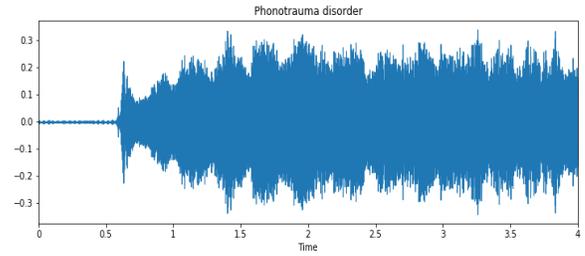

**Fig. 5** Waveform of Phonotrauma Voice disorder

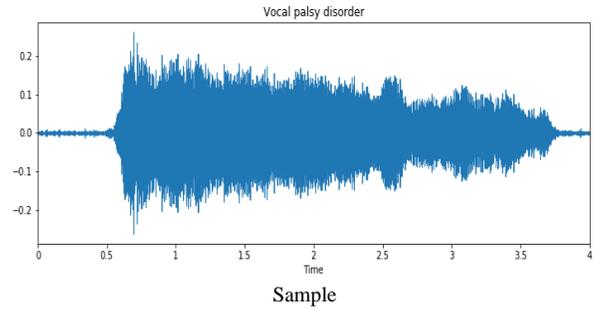

Sample

**Fig. 6** Waveform of Vocal Palsy Voice disorder

Data preprocessing includes feature extraction using different methods such as Mel-frequency cepstral coefficients (MFCC), spectral centroid, chroma and spectral contrast etc. For each voice sample we extracted 33 features combining all the feature extraction techniques. A brief overview of various feature extraction techniques are provided in below sections.

*a) Mel Frequency cepstral Coefficients (MFCC):* MFCC features are widely used in music genre classification, audio classification and speech recognition tasks, so we used it in this work. We extracted 13 MFCC features from each voice sample. More details on extracting MFCC features can be found at [18].

*b) Spectral Centroid:* Spectral centroid provides the center of mass of the spectrum. It provides the average loudness in terms of audio processing. One feature is extracted from each audio sample using spectral centroid. More details can be found at [19].

*c) Chroma:* Chroma provides a chromagram from a waveform. 12 features are extracted using chroma from each audio sample. More details can be found at [20].

*d) Spectral Contrast:* Spectral contrast represents the spectral characteristics of audio sample. We extracted 13 features using spectral contrast from each audio sample. More details can be found at [13].

## III. EXPERIMENTS AND RESULTS

This section provides the experiments and results to evaluate the effectiveness of our approach. The design of our LSTM network is shown in Table I showing one input layer with all 33 features extracted from each voice sample, 2 hidden layers out of which first hidden layer has 128 neurons while second one has 32 neurons and one output layer for predicting whether the voice sample is normal or having a voice disorder. We used the same experimental setup as [14] which was used for music genre classification as it provided promising results.

**Table I**: Design of LSTM Network

| Input Layer | 33 Input features extracted from audio samples |
|---|---|
| Hidden Layer I | 128 neurons |
| Hidden Layer II | 32 neurons |
| Output Layer | 4 outputs corresponding to 3 different voice disorders and 1 normal voice |

For training the LSTM model, optimizer used is Adams [15] while different batch sizes and epochs are used to get the best results. Categorical cross entropy is used as a loss function to measure the performance of classification model at each epoch. Increasing the number of epochs helps in improving the performance of the model.

**Table II**: Results of two phases

| Result Phase | Sensitivity | Specificity | UAR |
|---|---|---|---|
| I | 30% | 95.7% | 54% |
| II | 22% | 97.1% | 56% |

Table II shows the results obtained in two phases of results in FEMH Big data cup challenge. As we can see specificity is high in both the phases but sensitivity is low. Sensitivity determines the true positive rate while specificity true negative rate. Results show that normal voice people are correctly identified as normal as compared to the people with the voice disorder but unweighted average recall (UAR) represents the mean of recalls for both classes which increases with the number of epochs. In phase I we run the experiment for 500 epochs but in Phase II we run it for 5000 epochs.

Our results show that our approach works fine but requires more optimization in the future for better results.

## IV. CONCLUSION

This study presents a long short term memory (LSTM) approach to detect pathological voice disorders. The results show that it works fine in detecting the disorders. Also, different feature extraction techniques shows that these features can be beneficial for voice disorder detection. Future work includes more experiments with different hyperparameters to improve the results and use other feature extraction techniques too for further improvement.